\documentstyle[preprint,aps]{revtex}
\begin{document}

{\tighten
\preprint{\vbox{\hbox{CALT-68-1982}
                \hbox{hep-ph/9503209}
		\hbox{\footnotesize DOE RESEARCH AND}
		\hbox{\footnotesize DEVELOPMENT REPORT} }}

\title{Renormalons and confinement\footnote{%
Work supported in part by the U.S.\ Dept.\ of Energy under Grant no.\
DE-FG03-92-ER~40701.} }

\author{Ugo Aglietti$^{\,a,b,}$\footnote{Supported in part by the INFN.}
  and Zoltan Ligeti$^{\,a}$}

\address{ \vbox{\vskip 0.truecm}
  $^a$California Institute of Technology, Pasadena, CA 91125 \\
  \vbox{\vskip 0.truecm}
  $^b$Instituto Nazionale di Fisica Nucleare, 00185 Rome, Italy}

\maketitle

\begin{abstract}
We compute the renormalon ambiguity of the static potential, in the limit of a
large number of flavors.  An extrapolation of the QED result to QCD implies
that the large distance behavior of the quark potential is arbitrary in
perturbation theory, as there are an infinite number of prescriptions to
assign.  The shape of the potential at large distances is not only affected by
the renormalon pole closest to the origin of the Borel plane, but a resummation
of all renormalon contributions is required.  In particular, confinement can be
accommodated, but it is not explained.  At short distances there is no
indication of a linear term in the potential.
\end{abstract}

}%end tighten

\newpage

A way to learn about intrinsic limitations of perturbation theory in the
determination of a physical quantity is via the study of the large order
behavior of the perturbative expansion.  Although the coupling constant depends
only logarithmically on the scale, the perturbative expansion can produce power
suppressed corrections, as it is an asymptotic series.  Some nonperturbative
features of the theory can be inferred from the positions and residues of the
singularities in the Borel transform, with respect to the coupling constant
\cite{tHooft}.

At present it is not known how to compute renormalons in full QCD.  Most
calculations have been performed by summing fermion bubble insertions in the
gluon propagator.  The neglected graphs are presumably just as important as
those taken into account, so the result of summing the fermion bubble chain
does not prove the existence of renormalons in QCD.  Calculating renormalons
via bubble summation is rigorous in the limit of QED with a large number of
fermions, $N_f\to\infty$.  In this case a fermion bubble insertion in a photon
line yields a factor of $a=\alpha\,N_f$, which can be summed to all orders.
The contribution of other diagrams is suppressed by powers of $1/N_f$.
Extrapolating the large $N_f$ QED result to QCD is accomplished by assuming
that the ultraviolet and infrared properties can be interchanged.  Such studies
have received considerable attention in the literature recently, mostly in
connection with the heavy quark expansion (see, for example
\cite{BSUV2,BB,MW,BBZ,LMS,NeSa,Ugo,latt}).

In this letter we study renormalon ambiguities in a static property of QCD,
{\it i.e.}, the potential between a heavy quark and an anti-quark.
Neglecting color factors, the leading order one gluon exchange diagram
contributes to the potential by the usual Coulomb term
\begin{equation}\label{onegluon}
V_0(r) = -16\pi^2\,\alpha \int\! {{\rm d}^3 k\over(2\pi)^3}\,
  {e^{i\vec k\cdot\vec r}\over\vec k^2} = -{g^2\over r}\,.
\end{equation}
Diagrams with multiple gluon exchange are suppressed by powers of $1/N_f$,
while the insertion of fermion bubbles in the gluon line involves the coupling
$a$, which is treated as a parameter of order unity.  Therefore, we have to
consider an arbitrary number of bubbles inserted into the one gluon exchange
diagram.

If the perturbative expansion of the potential is given by
\begin{equation}\label{pert}
V(r) = \sum_{n=0}^\infty V_n\, (b_0\,\alpha)^{n+1} \,,
\end{equation}
then the Borel transform is defined as
\begin{equation}\label{borel}
\widetilde V(r) = \sum_{n=0}^\infty {V_n\over n!}\, u^n \,.
\end{equation}
Here $u$ is the Borel parameter.  For convenience, we defined the Borel
transformation with respect to $b_0\,\alpha$, where $b_0$ is the first
coefficient of the $\beta$-function,
$\beta(\alpha)=\mu^2\,\partial\alpha/\partial\mu^2
=-b_0\,\alpha^2+{\cal O}(\alpha^3)$.  In QED $b_0=-N_f/3\pi$, while in
QCD $b_0=(11-\frac23N_f)/4\pi$ \cite{poli}.
The reason for studying the Borel transform $\widetilde V(r)$ is that it
may converge even if the perturbative series for $V(r)$ in Eq.~(\ref{pert})
is divergent.  In that case the inverse Borel transformation
\begin{equation}\label{invborel}
V(r) = \int_0^\infty {\rm d}u\, e^{-u/(b_0\,\alpha)}\, \widetilde V(r) \,,
\end{equation}
provides a good definition of $V(r)$.  However, if $V_n$ grows at least
factorially for large $n$, then $\widetilde V(r)$ can have singularities on
the integration contour.  In our case these singularities will be isolated
poles on the positive real axis that arise from the infrared region of Feynman
diagrams.  These so-called infrared renormalons yield ambiguities in the
reconstruction of $V(r)$ from $\widetilde V(r)$, as the result depends on
the regularization of the pole contributions.

The Borel transform of the static potential can be calculated directly, using
the Borel transform of the resummed gluon propagator \cite{Bene}.  This yields
\begin{eqnarray}
\widetilde V(r) &=& -{16\pi^2\over b_0}\, \bigg({e^C\over\mu^2}\bigg)^{-u}
  \int\! {{\rm d}^3 k\over(2\pi)^3}\,
  {e^{i\vec k\cdot\vec r}\over(\vec k^2)^{1+u}} \nonumber\\*[4pt]
&=& -{4\,e^{-Cu}\over b_0}\, {1\over r}\, (\mu\,r)^{2u}\,
  {\Gamma(\frac12+u)\, \Gamma(\frac12-u)\over\Gamma(2u+1)} \,,
\end{eqnarray}
where $\mu$ is the renormalization scale, and $C$ is a regularization
scheme dependent constant.  We work in the momentum subtraction scheme
in which $C=0$.  The singularities of $\widetilde V(r)$ are simple poles
along the positive real axis at half-integer values of $u$
\begin{equation}
u_k=k+\frac12 \,, \qquad k=0,1,2,\ldots \,.
\end{equation}
The residues of these poles are
\begin{equation}
{\rm Res}\, \bigg[ \widetilde V(r) \Big|_{u_k} \bigg] = {1\over r}\,
  {4\over b_0}\, (\mu\,r)^{2k+1}\, {(-1)^k\over(2k+1)!} \,.
\end{equation}

The physical potential is reconstructed by the inverse Borel transform
in Eq.~(\ref{invborel})
\begin{eqnarray}\label{last}
V(r) = -{1\over r}\,{4\over b_0} \int_0^\infty {\rm d}u\, (\Lambda\,r)^{2u}\,
  {\Gamma(\frac12+u)\, \Gamma(\frac12-u) \over\Gamma(2u+1)} \,,
\end{eqnarray}
where $\Lambda$ is the pole of the coupling (the dynamically generated scale),
and we have used the relation $\Lambda^2/\mu^2=e^{-1/(b_0\,\alpha[\mu^2])}$.
To carry out this integration, the contour has to be deformed away from
the renormalon poles.\footnote{These renormalon poles are not present
if we include an infrared regulator, like a gluon mass or a vacuum
expectation value for the $\langle G_{\mu\nu}\,G^{\mu\nu}\rangle$ operator.
However, we are interested in what can be learned strictly from perturbation
theory.}
Using the QCD $\beta$-function coefficient, we obtain the ambiguity in
the quark potential
\begin{equation}\label{ambig}
\Delta V(r) = {1\over r}\, {4\over b_0}\, \sum_{k=0}^\infty c_k\,
  (\Lambda\,r)^{2k+1}\, {(-1)^k\over(2k+1)!} \,.
\end{equation}
Here $c_k$ are arbitrary complex numbers of order unity, related to the
ambiguity of regularizing the $1/(u-u_k)$ poles of $\widetilde V(r)$ by
their principal value plus $c_k$ times the delta function $\delta(u-u_k)$.

Inserting the QCD $\beta$-function into Eq.~(\ref{last}) is equivalent to
replacing the coupling constant evaluated at the fixed subtraction scale in the
one gluon exchange diagram with the running coupling at the scale of the gluon
momentum, $\alpha(k^2)$.  Diagrammatically this corresponds to dressing the one
gluon exchange diagram with self-energy and vertex corrections computed in the
leading logarithmic approximation.

The crucial point is that the static potential is an unambiguous physical
quantity.  It can be calculated on the lattice as the expectation value of a
Wilson loop.  In view of the above ambiguity originating from the renormalon
calculation, the static potential can only be free of ambiguities if
nonperturbative effects cancel Eq.~(\ref{ambig}).  It is in this sense how
renormalons `trace' nonperturbative effects.\footnote{There may also be
instanton contributions, and nonperturbative effects that do not correspond to
any local operator.}

There are a number of points to be made regarding our main result in
Eq.~(\ref{ambig}):

$a.$  In previous calculations, powers of some large mass scale or momentum
flow suppressed the renormalon contributions of the poles further away from
the origin of the Borel plane.  In our case this scale is replaced by $1/r$.
At large distances $\Lambda/(1/r)$ enhances the contributions of the poles far
from the origin.  Therefore, the ambiguity in the static potential at large
distances is affected by all renormalon poles.

$b.$  By varying the values of $c_k$, the renormalon ambiguity at large
distances becomes arbitrary.  Taking for example $c_k=1$ yields
\begin{equation}
\Delta V(r) \sim {\sin(\Lambda\,r)\over r} \,,
\end{equation}
while $c_k=(-1)^k$ yields
\begin{equation}
\Delta V(r) \sim {\sinh(\Lambda\,r)\over r} \,.
\end{equation}
In the first case, the renormalon calculation implies that nonperturbative
correction to the potential is not confining, while it is confining in the
latter case.  With a proper choice of the $c_k$'s, an arbitrary potential
shape can be produced at large distances.  For example, the $c_k$'s can
reproduce a nonperturbative correction to the potential of the form
$\Lambda\,\sqrt{1+(\Lambda\,r)^2}$, {\it i.e}, a linear behavior
in $r$ for $r\gg\Lambda^{-1}$.

$c.$  The renormalon ambiguities do not affect the Coulomb part of the
potential, only terms that are proportional to $r^{2k}$.  Near $r=0$ the
renormalon calculation gives no indication of a linearly rising correction to
the potential.  It seems conceivable to us that this is a real feature of QCD,
namely `flux tubes' develop only for $r>\Lambda^{-1}$.  Our result does not
constrain the large $r$ behavior: as was shown above, $V(r)$ can even grow
linearly for large $r$.

$d.$  There is an analogy with non-renormalizable theories, where knowledge of
an infinite number of counterterms is needed to derive scattering amplitudes
from the Lagrangian.  In the present case, the values of $c_k$ would specify
the gluon propagator at large distances.

These observations imply that the renormalon calculation supplemented with
bubble summation has a very limited capability in describing long distance QCD.
In this sense our results question the suitability of resumming fermion bubble
chains for phenomenological estimates of nonperturbative effects in real QCD.
One could argue that some values of $c_k$ imply confinement, but based on the
information available from the renormalon calculation, such a choice is
completely arbitrary, and is not any better motivated than values of $c_k$ that
suggest that nonperturbative contributions to the static potential need not be
confining.  In this framework, therefore, the Landau pole gives no indication
of what type of nonperturbative effects occur.  Beyond the perturbatively
calculable result there are an infinite number of parameters that need to be
specified to describe the long distance dynamics.

\acknowledgements
We thank Guido Martinelli, John Preskill, and especially Mark Wise
for discussions.

{\tighten

} %end tighten (references & figure captions)

\end{document}